\documentstyle [12pt] {article}

\catcode`@=12
\newcommand{\be}{\begin{equation}}
\newcommand{\ee}{\end{equation}}
\newcommand{\ber}{\begin{eqnarray}}
\newcommand{\eer}{\end{eqnarray}}
\newcommand{\bers}{\begin{eqnarray*}}
\newcommand{\eers}{\end{eqnarray*}}
\begin{document}
\vspace{0.5in}
\oddsidemargin -.375in
\def\ee{\end{equation}}
\thispagestyle{empty}
\begin{flushright} ISU-NP-98-04\\
IITAP-98-002\\
July 1998\
\end{flushright}
\vspace {.5in}
\begin{center}
{\Large\bf $B_c$ mesons in a Bethe-Salpeter model\\}
\vspace{.5in}
{\bf A. Abd El-Hady${}^a$, 
M.A.K. Lodhi ${}^b$ and J.~P.~Vary ${}^c$\\}
\vspace{.1in}
${}^{a)}$  
{\it
Physics Department, Zagazig University, Zagazig, Egypt}\\
\vspace{.1in}
${}^{b)}$  
{\it
Department of Physics, Texas Tech University, Lubbock, Texas 79409, USA} \\ 
\vspace{.1in}
${}^{c)}$  
{\it
 Department of Physics and Astronomy, Iowa State University , Ames, Iowa
50011, USA}\\
\vskip .5in
\end{center}

\vskip .1in \begin{abstract} 

We apply our Bethe-Salpeter model for mesons to the $B_c$ family with
parameters fixed in our previous investigation. We evaluate the mass of
the pseudo-scalar $B_c$ meson as 6.356 GeV/$c^2$ and 6.380 GeV/$c^2$ and
the lifetime as 0.47 ps and 0.46 ps respectively in two reductions of
the Bethe-Salpeter Equation, in good agreement with the recently reported
mass of 6.40 $\pm$ 0.39 (stat.) $\pm$ 0.13 (syst.) GeV/$c^2$ and
lifetime of $0.46^{+0.18}_{-0.16}$ (stat.) $\pm$ 0.03 (syst.) ps by the
CDF Collaboration. We evaluate the decay constant of the $B_c$ meson and
compare different contributions to its decay width.

\end{abstract}
\vskip .25in
\newpage

%
%

Recently the CDF Collaboration reported the observation of the
bottom-charmed mesons $B_c$\ in 1.8~TeV $p\overline{p}$ collisions using
the CDF detector at the Fermilab Tevatron \cite{CDF}. This pseudo-scalar
state is the lowest energy state of the family of mesons composed of a
$\bar b$ anti-quark and a c quark. Since this state lies below the ($B
D$) threshold and has non-vanishing flavor quantum numbers, it decays
only through weak interactions. This eliminates uncertainties
encountered in strong decays and renders the decay width of $B_c$ more
theoretically tractable.

Different approaches have been used to evaluate the spectrum of $B_c$
mesons. Non-relativistic potential models have been used by Eichten and
Quigg \cite{EQ} where they discussed four potentials and gave mass
values for the $B_c$ meson in the range 6.248--6.266 GeV/$c^2$.
Gershtein {\it et al.} \cite{GER} used two potentials and reported
predictions of 6.253 and 6.264. QCD sum rules have been used by Chabab
\cite{chabab} where he predicted a prediction of 6.25 GeV/$c^2$.

In this paper, we extend our model \cite{sommerer2, model} based on the
Bethe-Salpeter Equation (BSE) to include the bottom-charmed mesons. BSE
provides an appealing starting point to describe hadrons as relativistic
bound states of quarks, just as the Dirac Equation provides a
relativistic description of a fermion in an external field.  The BSE for
a bound state may be written in momentum space in the form
\begin{eqnarray}
G^{-1}(P,p)\psi(P,p)=\int\frac{1}{(2\pi)^{4}}V(P,p-p')\psi(P,p')d^4p'
\end{eqnarray}  
Where $P$ is the four-momentum of the bound state, $p$ is the relative
four-momentum of the constituents . The BSE has three elements, the
interaction kernel ($V$) and the propagator ($G$) which we provide as
input, and the amplitude ($\psi$) obtained by solving the equation. We
also solve for the energy, which is contained in the
propagator.

Different approaches have been developed to make the four dimensional
problem BSE more tractable and physically appealing. These include the
Instantaneous Approximation (IA) and Quasi-Potential Equations (QPE)
\cite{itzykson,brown,silvestre}. In the IA, the interaction kernel is
taken to be independent of the relative energy. In QPE, the two particle
propagator is modified in a way which keeps covariance and reduces the
4-dimensional BSE to a 3-dimensional equation. Of course, there is
considerable freedom in carrying out this reduction.

We have used two reductions of the QPE to study the meson spectrum
\cite{sommerer2, model}. These reductions correspond to different choices of the
two particle propagator used to reduce the problem into three
dimensions. We refer to these reductions as A and B.  Reduction A
corresponds to a spinor form of the Thompson equation \cite{Thompson}
and reduction B corresponds to a new QPE introduced in
Ref. \cite{sommerer}. These two reductions are chosen because they are
shown to give good fits to the meson spectrum.

We assume the interaction kernel to consists of
a one gluon exchange interaction, $V_{OGE}$, in the ladder
approximation, and a phenomenological, long range scalar confinement
potential, $V_{CON}$ given in the form
\begin{eqnarray}
V_{OGE}+V_{CON} & = & -{4\over
3}\alpha_s{\gamma_\mu\otimes\gamma_\mu\over {(p-p')^2}}
+\sigma{\rm\lim_{\mu\to 0}}{\partial^2\over\partial\mu^2} {{\bf
1}\otimes{\bf 1}\over-(p-p')^2+\mu^2} \ 
\end{eqnarray}
Here, $\alpha_s$ is the strong coupling, which is weighted by the meson
color factor of ${4\over 3}$, and the string tension $\sigma$ is the
strength of the confining part of the interaction.  We adopt a scalar
Lorentz structure $V_{CON}$ as discussed in \cite {sommerer2, model}.

In our model the strong coupling is assumed to run as in the leading log
expression for $\alpha_s$,
\begin{eqnarray} \alpha_s(Q^2) & = & {4\pi\alpha_s(\mu^2)\over
4\pi+\beta_1\alpha_s(\mu^2){\rm ln}\bigl({Q^2/\mu^2}\bigr)} \
\end{eqnarray} 
where $\beta_1=11-2n_f/3$ and $n_f$= 4 is the number of quark flavors
taken to be fixed. At the scale of the Z-boson,
$\alpha_s(\mu^2=M_Z^2)\simeq 0.12$ and $Q^2$ is related to the meson
mass scale through,
\begin{equation}
Q^2 = \gamma^2 M_{meson}^2 + \beta^2,
\end{equation}
where $\gamma$ and $\beta$ are parameters determined by a fit to the
meson spectrum. In our formulation of BSE there are therefore seven
parameters : four masses, $m_{u}$=$m_{d}$, $m_s$, $m_{c}$, $m_{b}$; the
string tension $\sigma$, and the parameters $\gamma$ and $\beta$ used to
govern the running of the coupling constant.

\begin{table}[h!tb]
\caption{Values of the parameters used in reductions A, B}
\begin{center}
\begin{tabular}{|c|c|c|}
\hline
   &Reduction A   &  Reduction B  \\
\hline
$m_b$ (GeV) &4.65&4.68\\
$m_c$ (GeV) &1.37&1.39\\
$m_s$ (GeV) &0.397&0.405\\
$m_u$ (GeV) &0.339&0.346\\
$\sigma$ (GeV$^2$) &0.233&0.211\\
$\gamma$  &0.616&0.444\\
$\beta$ (GeV)  &0.198&0.187\\
\hline
\end{tabular}\\
\label{tp71}
\end{center}
\end{table}

In Ref. \cite{sommerer2, model} we fitted the meson spectrum using these
seven parameters. However we did not include the $B_c$ mesons in our
fit. Table \ref{tp71} shows the values of the parameters obtained by the
fits in reductions A and B.

In this paper we extend our model to evaluate the properties of the
$B_c$ mesons using these same values of the parameters. In Table
\ref{tp72} we compare the spectrum of $B_c$ mesons obtained in reduction A and B
with the work of Eichten and Quigg \cite{EQ}, Gershtein at al \cite{GER}
using Martin potential and Gershtein {\it et al.} \cite{GER} using
Buchmuller-Tye (BT) potential. The first row compares with the
experimental results \cite{CDF} of 6.40 $\pm$ 0.39 (stat.) $\pm$ 0.13 (syst.)
GeV/$c^2$. Both reduction A and B compare reasonably well with the
experimental results though the experimental uncertainties are large.
\begin{table}[h!tb]
\caption{spectrum of $B_c$ mesons in different channels (GeV/$c^2)$.}
\begin{center}
\begin{tabular}{|c|c|c|c|c|c|c|}
\hline

State & This work   & This work   & Eichten and Quigg &Gershtein {\it et al.}\cite{GER}&
Gershtein {\it et al.}\cite{GER}\\
      & Reduction A & Reduction B & Ref. \cite{EQ} & Martin potential               &
BT potential                     \\
\hline
$1^1S_0$& 6.356 & 6.380 & 6.264 & 6.253 & 6.246 \\
$1^3S_1$& 6.397 & 6.415 & 6.337 & 6.317 & 6.337 \\
$1^3P_0$& 6.673 & 6.692 & 6.700 & 6.683 & 6.700 \\
$1^3P_2$& 6.751 & 6.773 & 6.747 & 6.743 & 6.747 \\
$1^1P_1$& 6.752 & 6.777 &       & 6.729 & 6.736 \\
$2^1S_0$& 6.888 & 6.874 & 6.856 & 6.867 & 6.856 \\
$2^3S_1$& 6.910 & 6.891 & 6.899 & 6.902 & 6.899 \\
$1^3D_1$& 6.984 & 6.955 & 7.012 & 7.008 & 7.012 \\
\hline
\end{tabular}
\label{tp72}
\end{center} 
\end{table}

In our formalism the mesons are taken as bound states of a quark and an
        anti-quark. The wavefunctions for the mesons are calculated by
        solving reductions of Bethe-Salpeter equation \cite {sommerer2,
        model}.  We construct the meson states as \cite{ISGW1}
\begin{eqnarray}
|M({\bf {P_M}},J,m_J)\rangle\  & = &
\sqrt{2M} \int d^{3}{\bf p} \langle L m_{L}S m_{S}|J m_J\rangle\  \langle
s m_s \bar{s} m_{\bar{s}}|S m_S\rangle\ \nonumber\\
 & &\Phi_{L m_L}({\bf p})|\bar q( {m_{\bar
q} \over \ M_{q \bar q} } {\bf {P}_M} - {\bf {p}},m_{\bar s})
\rangle|q({m_q \over \ M_{q \bar q} } {\bf {P}_M} + {\bf {p}},m_s)\rangle
\label{states}
\end{eqnarray}
where the quark states are given by 
\begin{eqnarray}
|q({\bf {p}},m_s)\rangle\ &=& \sqrt{\frac{(E_q + m_q)}{2m_q}} \pmatrix{
 \chi^{m_s} \cr \ {{\bf{\sigma}}\cdot{\bf p }\over
{(E_q+m_q)}}  \chi^{m_s} \cr }\nonumber\\
M_{q \bar q}&=&m_q+m_{\bar q}\nonumber\\ 
E_q&=&\sqrt{m^2_{q}+{\bf{p}}^2} 
\end{eqnarray}
 
In Eq. \ref{states} $M$ is the meson mass.  The meson and the
constituent quark states satisfy the normalization condition.
\begin{eqnarray} \langle
M({\bf{P^\prime}_M},J^{\prime},m^{\prime}_J)|M({\bf {P}_M},J,m_J)
\rangle\ &=& 2E\delta^3({\bf
{P^\prime}_M}-{\bf{P}_M})\delta_{J^{\prime},J}\delta_{m^{\prime}_J,m_J}
\end{eqnarray} \begin{eqnarray} \langle
q({\bf{p^\prime}},m^{\prime}_s)|q({\bf{p}},m_s) \rangle\ &=&
{E_q\over{m_q}} \delta^3({\bf {p^\prime}}-{\bf
{p}})\delta_{m^{\prime}_s,m_s} \end{eqnarray}

In previous works \cite{BD1,BD2,BD3}, we have used the
wavefunctions of our model to evaluate the semi-leptonic form factors
for $B$ to $D$ and $D^*$ mesons, and the leptonic decay
constants. Here, we are interested in the leptonic decay constant. The
weak decay constants for the pseudo-scalar and vector mesons are defined
by
\ber
<0|J_{\mu}|P(p)> & = & i f_P p_{\mu}\nonumber\\
<0|J_{\mu}|V(p)> & = & M_Vf_V \varepsilon_{\mu}\nonumber\\
J_{\mu} & = & V_{\mu} -A_{\mu}\
\eer
where $P$ and $V$ are pseudo-scalar and vector states and $V_{\mu}$ and
$A_{\mu}$ are the vector and axial vector currents.

Taking into account the relativistic effects, the expressions of the
decay constants in terms of the wavefunctions are given by \cite{VI}
\ber
f_i & = & \sqrt{\frac{12}{M}}
\int^{\infty}_{0}\frac{p^2dp}{2 \pi^3}
\sqrt{\frac{(m_q + E_q)(m_{\bar{q}} + E_{\bar q})}{4E_qE_{\bar{q}}}}
F_i(p)\\
\eer
where the subscript $i$ represents $P$ or $V$ and 
\ber
F_{P}(p) & = & \left[ 1-\frac{p^2}{
(m_q + E_q)(m_{\bar{q}} + E_{\bar q})}\right]\psi_P(p)\\
F_V(p) & = & \left[ 1-\frac{p^2}{
3(m_q + E_q)(m_{\bar{q}} + E_{\bar q})}\right]\psi_V(p)\
\eer
where $\psi_{P}$, $\psi_{V}$ are the wavefunctions of the pseudo-scalar and
vector states respectively. The non-relativistic limit of these
expressions yields a relation between $f_{i}$ and the wavefunction at the origin in 
coordinate space , $R(0)$,
\ber
f_i=\sqrt{\frac{3}{\pi M}}R(0).
\eer

The leptonic decay constant $(f_{B_c})$ is relevant for the annihilation
channel of the $B_c$ pseudo-scalar meson. In Table \ref{tp73} we compare
different predictions for this quantity.

\begin{table}[h!tb]
\caption{
Leptonic decay constant of $B_c$ $(f_{B_c})$ in MeV.
}
\begin{center}
\begin{tabular}{|c|c|c|c|c|c|}
\hline
This work & This work & Eichten and Quigg & Gershtein {\it et al.} \cite{GER}         &  Gershtein {\it et al.} \cite{GER}\\
A         & B         & Ref. \cite{EQ} & Martin potential  &  BT potential\\
\hline
578       & 490       & $500$&$512 $        & $500$\\
\hline
\end{tabular}
\label{tp73}
\end{center}
\end{table}

The lifetime of $B_c$ is a very important quantity which may help us
understand the basic properties of the weak interaction at a fundamental
level especially since the strong interaction effects can be estimated
reliably.  The total width can be approximated by the sum of the widths
of $\bar b$-quark decay with the spectator $c$-quark, the $c$-quark
decay with the spectator $\bar b$-quark, and the annihilation channel
$B_c^+\rightarrow l^+\nu_l (c\bar s, u\bar s)$, $l=e,\; \mu,\; \tau$.
Since all these decays lead to different final states, we have no
interference between different amplitudes. The total width is then given
by
\begin{equation}
\Gamma (B_c\rightarrow X)=\Gamma (b\rightarrow X)
+\Gamma (c\rightarrow X)+\Gamma \mbox{(ann)}\;.
\end{equation}

Neglecting the quark binding effects, we obtain for the $b$ and $c$
inclusive widths in the spectator approximation,
\begin{eqnarray}
\Gamma (b\rightarrow X) & = & \frac{G^2_F|V_{cb}|^2m^5_b}{192\pi^3}\cdot 9 \;,
 \nonumber \\
\Gamma (c\rightarrow X) & = & \frac{G^2_F|V_{cs}|^2m^5_c}{192\pi^3}\cdot 5  \;,
\label{d2}
\end{eqnarray}
The width of the annihilation channel is given by
\begin{equation}
\Gamma \mbox{(ann)} =\sum_i\frac{G^2_F}{8\pi}
|V_{bc}|^2f^2_{B_c}M_{bc} m^2_i \biggl(1-\frac{m^2_i}{m^2_{Bc}}\biggr)^2\cdot
C_i\;,
\label{d3}
\end{equation}
where $C_i = 1$ for the $\tau\nu_\tau$ channel and $C_i =3|V_{cs}|^2 $
for the $\bar c s$ channel, and $m_i$ is the mass of the heaviest
fermion ($\tau$ or $c$). Table \ref{tp74} shows various contributions to
the width of $B_c$ in our model.
\begin{table}[h!tb]
\caption{
Various contributions to the decay width of $B_c$ in $10^{-12}$ GeV.
}
\begin{center}
\begin{tabular}{|c|c|c|c|}
\hline
& $\Gamma (b \rightarrow$ X)&$\Gamma (c \rightarrow$ X)&$\Gamma$ (ann)   \\
\hline
Reduction A&0.75&0.51&0.14\\
Reduction B&0.78&0.55&0.11\\
\hline
\end{tabular}\\
\label{tp74}
\end{center}
\end{table}

We have used $V_{cb}=0.041$, and $V_{cs}=0.96$. From Table \ref{tp74} we
see that both reductions predict that the $b$ decay dominates $c$ decay in
$B_c$ meson.

In Table \ref{tp75}, we compare the lifetime of $B_c$ in different
models with the CDF experimental result. The experimental result
indicates that the binding effects may not be very important as suggested
by Quigg \cite{Quigg}. 
\begin{table}[h!tb]
\caption{Comparison of the lifetime of $B_c$ meson (in ps) in different models.}
\begin{center}
\begin{tabular}{|c|c|c|c|c|}
\hline
Experiment \cite{CDF} & Reduction A & Reduction B & Quigg \cite{Quigg}& Gershtein {\it et al.} \cite{GER}\\
\hline
$0.46^{+0.18}_{-0.16}$ (stat.) $\pm$ 0.03 (syst.) & 0.47 &0.46 &  $1.1-1.4$&$0.55 \pm0.15 $\\
\hline
\end{tabular}
\label{tp75}
\end{center}
\end{table}

In conclusion, we have evaluated the meson spectrum of the $B_c$ mesons
in two reductions of BSE. We used parameters fixed from our previous
fits and our results for properties of $B_c$ agree with the recent
measurement of the CDF Collaboration of $B_c$ mass. We also predicted
the leptonic decay constant and evaluated various contributions to the
decay width of $B_c$. The partial width of $B_c$ due to b-quark decay
dominates that due to the c-quark decay. Our result for the $B_c$
lifetime is in good agreement with the CDF measurement.

{\bf Acknowledgments}

This work was supported in part by the US Department of Energy, Grant
No. DE-FG02-87ER40371, Division of High Energy and Nuclear Physics and
by the International Institute of Theoretical and Applied Physics, Ames,
Iowa.


\newpage

\begin{thebibliography}{References}


\bibitem{CDF}
F. Abe {\it et al.} The CDF Collaboration, hep-ex/9804014; W.Hav and V.
Papadimitriou, private communication.


\bibitem{EQ}
E. Eichten and C. Quigg, Phys. Rev. {\bf D49}, 5845 (1994). 


\bibitem{GER}
S.S. Gershtein, V.V. Kiselev, A.K. Likhoded, A.V. Tkabladze,
A.V. Berezhnoi, A.I.  Onishchenko, hep-ph/9803433; V.V. Kiselev,
A.K. Likhoded, A.V. Tkabladze, Phys. Rev. {\bf D51}, 3613 (1995);
S.S. Gershtein, V.V. Kiselev, A.K. Likhoded, A.V. Tkabladze,
Phys. Usp. {\bf 38}, 1 (1995).


\bibitem{chabab}
M. Chabab, LMPHE/97-04; Phys. Lett. {\bf B325}, 205 (1994).  


\bibitem{sommerer2}
A. J. Sommerer, J. R. Spence, and J. P. Vary, Phys. Rev. {\bf C 49}, 513
(1994).


\bibitem{model}
Alan J. Sommerer, A. Abd El-Hady, John R. Spence, and James P. Vary, 
Phys. Lett. {\bf B348}, 277 (1995). 


\bibitem{itzykson}C. Itzykson, and J.B. Zuber, Quantum Field theory.
McGraw-Hill, New York (1980) (Chapter 10 gives a review of
Bethe-Salpeter equation).

\bibitem{brown}G. E. Brown and A. D. Jackson, The Nucleon-Nucleon
Interaction North-Holland, New York (1976); and references therein.

\bibitem{silvestre}B. Silvestre-Brac, A. Bilal, C. Gignoux, and P.
Schuck, Phys. Rev. D {\bf 29}, 2275 (1984).




\bibitem{Thompson}R. H. Thompson, Phys. Rev. {\bf D 1}, 110 (1970).

\bibitem{sommerer}J. R. Spence and J. P. Vary, Phys. Rev. {\bf C 47}, 1282
(1993); A. J. Sommerer, J. R. Spence, and J. P. Vary, Mod.
Phys. Lett. {\bf A 8}, 3537 (1993).




\bibitem{ISGW1}
Nathan Isgur, Daryl Scora, Benjamin Grinstein, and Mark B. Wise, Phys. Rev. 
{\bf D 39}, 799 (1989).


\bibitem{BD1}
A. Abd El-Hady, K.S. Gupta, A.J. Sommerer, J. Spence, and  J.P. Vary, 
Phys. Rev. {\bf D51}, 5245 (1995). 

\bibitem{BD2}
A. Abd El-Hady, A. Datta, K.S. Gupta, and  J.P. Vary, 
Phys. Rev. {\bf D55}, 6780 (1997). 

\bibitem{BD3}
A. Abd El-Hady, A. Datta, and  J.P. Vary, 
Phys. Rev. {\bf D58}, 6780 (1998). 


\bibitem{VI} S. Veseli and I. Dunietz Phys. Rev. {\bf D 54} 6803 (1996).

\bibitem{Quigg}
C. Quigg, Proceedings of the Workshop on B Physics at Hadron Accelerators, ed.
  by P. McBride and C. Shekhar Mishra, Fermilab-CONF-93/267 (SSCL-SR-1225)
  (1994).



\end{thebibliography}
\end{document}